\documentclass[12pt,english,sort&compress]{article}
\usepackage[T1]{fontenc}
\usepackage[latin9]{inputenc}
\usepackage{fancyhdr}
\usepackage{babel}
\usepackage{booktabs}
\usepackage{varwidth}
\usepackage{amsmath}
\usepackage{graphicx}
\usepackage[a4paper]{geometry}
\usepackage{setspace}
\usepackage[numbers]{natbib}
\usepackage[pdfusetitle,
 bookmarks=true,bookmarksnumbered=false,bookmarksopen=false,
 breaklinks=false,pdfborder={0 0 0},pdfborderstyle={},backref=false,colorlinks=false]
 {hyperref}

\makeatletter

\providecommand{\tabularnewline}{\\}
\newenvironment{cellvarwidth}[1][t]
    {\begin{varwidth}[#1]{\linewidth}}
    {\@finalstrut\@arstrutbox\end{varwidth}}

\usepackage{times}

\date{}

\renewcommand{\@openbib@code}{\setlength{\itemsep}{-1pt}}

\renewcommand{\subsectionmark}[1]{}

\usepackage[compact]{titlesec}
\titleformat{\section}{\LARGE \bfseries}{\thesection}{1em}{}
\titleformat{\subsection}{\large \bfseries}{\thesubsection}{1em}{}

\makeatother

\begin{document}
\global\long\def\d{\mathrm{d}}%
\global\long\def\boltzmann{k_{b}}%
\global\long\def\chains{N_{0}}%

\global\long\def\Tvptt{T_{VPTT}}%

\global\long\def\xh{\hat{\mathbf{x}}}%
\global\long\def\yh{\hat{\mathbf{y}}}%
\global\long\def\zh{\hat{\mathbf{z}}}%

\global\long\def\defgradTg{\mathbf{F}_{g}}%
\global\long\def\stretchg{\lambda_{g}}%

\global\long\def\defgradTs{\mathbf{F}_{s}}%
\global\long\def\stretchs{\lambda_{s}}%
\global\long\def\defgradTsIso{\tilde{\mathbf{F}}_{s}}%

\global\long\def\defgradT{\mathbf{F}}%
\global\long\def\stretch{\lambda}%

\global\long\def\n{n}%
\global\long\def\L{L}%
\global\long\def\rat{\rho}%
\global\long\def\freeC{\psi_{c}}%
\global\long\def\entropyC{S_{c}}%
\global\long\def\forceC{f_{c}}%
 
\global\long\def\etoe{r}%

\global\long\def\fcr{\tilde{f}}%

\global\long\def\nmin{n_{g}}%
\global\long\def\Lmin{L_{g}}%
\global\long\def\nmax{n_{s}}%
\global\long\def\Lmax{L_{s}}%
\global\long\def\lang{\beta}%
\global\long\def\ratm{\rho_{m}}%
\global\long\def\ratmin{\rho_{1}}%
\global\long\def\ratmax{\rho_{2}}%

\global\long\def\refptsT{\mathbf{x}}%
\global\long\def\curptsT{\mathbf{y}}%
\global\long\def\RT{\mathbf{R}}%
\global\long\def\R{R}%
\global\long\def\RhT{\hat{\mathbf{R}}}%
\global\long\def\rT{\mathbf{r}}%
\global\long\def\r{r}%
\global\long\def\rhT{\hat{\mathbf{r}}}%
\global\long\def\Js{J_{s}}%

\global\long\def\free{\psi}%
\global\long\def\entropy{S}%
\global\long\def\freen{\psi_{n}}%
\global\long\def\freem{\psi_{m}}%
\global\long\def\cp{c_{p}}%
\global\long\def\water{m}%

\global\long\def\prs{p_{s}}%
\global\long\def\prg{p_{g}}%
\global\long\def\forces{f_{s}}%
\global\long\def\forceg{f_{g}}%
\global\long\def\chempot{\mu}%
\global\long\def\volw{v_{w}}%

\global\long\def\stressTc{\boldsymbol{\sigma}_{c}}%
\global\long\def\stressTs{\boldsymbol{\sigma}_{s}}%
\global\long\def\stressTg{\boldsymbol{\sigma}_{g}}%

\global\long\def\timec{\tau}%
\global\long\def\stretchr{\lambda_{r}}%

\title{\vspace{-80pt}Phase coexistence in thermo-responsive PNIPAM hydrogels
triggered by mechanical forces}
\author{Noy Cohen $^{\footnotesize{a,}} \footnote{e-mail address: noyco@technion.ac.il}\,\,$
\\
$^{a}${\footnotesize{Department of Materials Science and Engineering, Technion  - Israel Institute of Technology, Haifa 3200003, Israel}}}
\maketitle
\begin{abstract}
Poly(N-isopropylacrylamide) (PNIPAM) is a temperature-responsive polymer
that undergoes large volumetric deformations through a transition
from a swollen to a collapsed state at the volume phase transition
temperature (VPTT). Locally, these deformations stem from the coil-to-globule
transition of individual chains. In this contribution, I revisit the
study of \citet{Suzuki&ishii99JCP}, which demonstrated that a PNIPAM
rod can exhibit phase coexistence (i.e. comprise swollen and collapsed
domains) near the VPTT when subjected to mechanical constraints. Specifically,
that paper showed that (1) collapsed domains gradually form in a fixed
swollen rod with time and (2) swollen domains can nucleate in a collapsed
rod that under uniaxial extension. These behaviors originate from
the local thermo-mechanical response of the chains, which transition
between states in response to the applied mechanical loading. Here,
I develop a statistical-mechanics based framework that captures the
behavior of individual chains below and above the VPTT and propose
a probabilistic model based on the local chain response that sheds
light on the underlying mechanisms governing phase nucleation and
growth. The model is validated through comparison with experimental
data. The findings from this work suggest that in addition to the
classical approaches, in which the VPTT is programmed through chemical
composition and network topology, the transition can be tuned by mechanical
constraints. Furthermore, the proposed framework offers a pathway
to actively tailor the VPTT through the exertion of mechanical forces,
enabling improved control and performance of PNIPAM hydrogels in modern
applications. 
\end{abstract}


\section{Introduction}

Poly(N-isopropylacrylamide) (PNIPAM) is a thermo-responsive macromolecule
that undergoes a local reversible coil-to-globule transition on the
chain level and a macroscopic swollen-to-collapsed volumetric deformation
at a volume phase transition temperature (VPTT) \citep{Afroze2000,Das&etal24AFM,brighenti&etal20FM}.
Typically, the VPTT is $\Tvptt\sim32-34^{\circ}C$, with variations
that depend on different factors such as composition \citep{jain&etal15PC,Suzuki&Sanda97JCP,Juurinen2014},
solvent type \citep{Winnik90Macromol,Kojima&tanaka12SM}, and the
application of external force \citep{Suzuki&Sanda97JCP,Suzuki&ishii99JCP}.
The volumetric deformations associated with the transition are significant,
with the literature reporting over a 10 times increase in volume from
the collapsed to the swollen state \citep{Suzuki&Sanda97JCP,Drozdov2015,shen&etal19SM,brighenti&etal20FM,drozdov&chris21JMBBM}. 

The sharp thermally-induced phase transition in PNIPAM hydrogels lends
itself to many applications. For example, PNIPAM hydrogels have been
proposed for tissue engineering \citep{nakayama&etal10MM,Wu2023,Rana2021}
and drug delivery systems \citep{guan11SoftMat,Ashraf2016,Throat2024},
in which the temperature-dependent response is exploited for cell
scaffolding and controlled release. In soft robotics, PNIPAM hydrogels
are used in the design of shape-memory materials \citep{Zhang2015},
smart switches \citep{Zhang2020}, and deformable structures that
exploit the temperature-dependent volume changes \citep{gao2018synergistic,zhou2016waveguiding}.

Interest in PNIPAM hydrogels has steadily increased over the past
decades. Beginning with the pioneering work of \citet{hirokawa&tanaka84AIP},
who demonstrated the sharp volume phase transition in non-ionic gels,
many experiments and modeling attempts have been carried out. At the
chain-level perspective, several studies focused on the local response
before and after the VPTT \citep{Liang&naka18,Levin2025,drozdov&chris21JMBBM,Pang&cui13Lang,Cui&etal12Lngm}.
The bulk response of PNIPAM gel was investigated through continuum-based
models \citep{Drozdov2015,Zheng2022,Drozdov2014,drozdov&chris21JMBBM,Levin2025,Chester2011,Chester2015},
as well as different experimental set-ups \citep{takigawa&etal98PGN,Suzuki&Sanda97JCP,Suzuki&ishii99JCP,shen&etal19SM}. 

The behavior of PNIPAM hydrogels has been traditionally characterized
by their response at states below and above the VPTT, with the transition
captured via changes in the phenomenological dimensional interaction
parameter $\chi$, as proposed by Flory \citep{Flory1942,Huggins1942}.
The work of \citet{Levin2025} established a fundamental framework
to capture the network response by focusing on how microstructural
quantities such as chain length and distribution evolve and govern
the equilibrium response, and accordingly described the transition
from collapsed to swollen and vice versa. However, a significant challenge
remains in understanding the microstructural evolution of the network
and the influence of mechanical forces around the transition region
itself. The insightful work of \citet{Suzuki&ishii99JCP} experimentally
demonstrated that around the VPTT an applied mechanical loading can
induce phase coexistence, i.e. the simultaneous and stable presence
of collapsed and swollen phases, as shown in Fig. \ref{fig:coexistence_demonstration}.
This contribution revisits that work by building on the findings of
\citet{Levin2025} and introducing a framework that delineates the
underlying mechanisms governing phase coexistence by explicitly accounting
for the stochasticity of conformational chain transitions and the
influence of external forces. This addition is critical to capturing
the nucleation and growth of new phases, as it provides a quantifiable
description of the transition regime and enables one to tune the VPTT
through mechanical forces. 

\subsubsection*{Phase coexistence in PNIPAM networks}

\begin{figure}
\hspace*{\fill}\includegraphics[width=12cm]{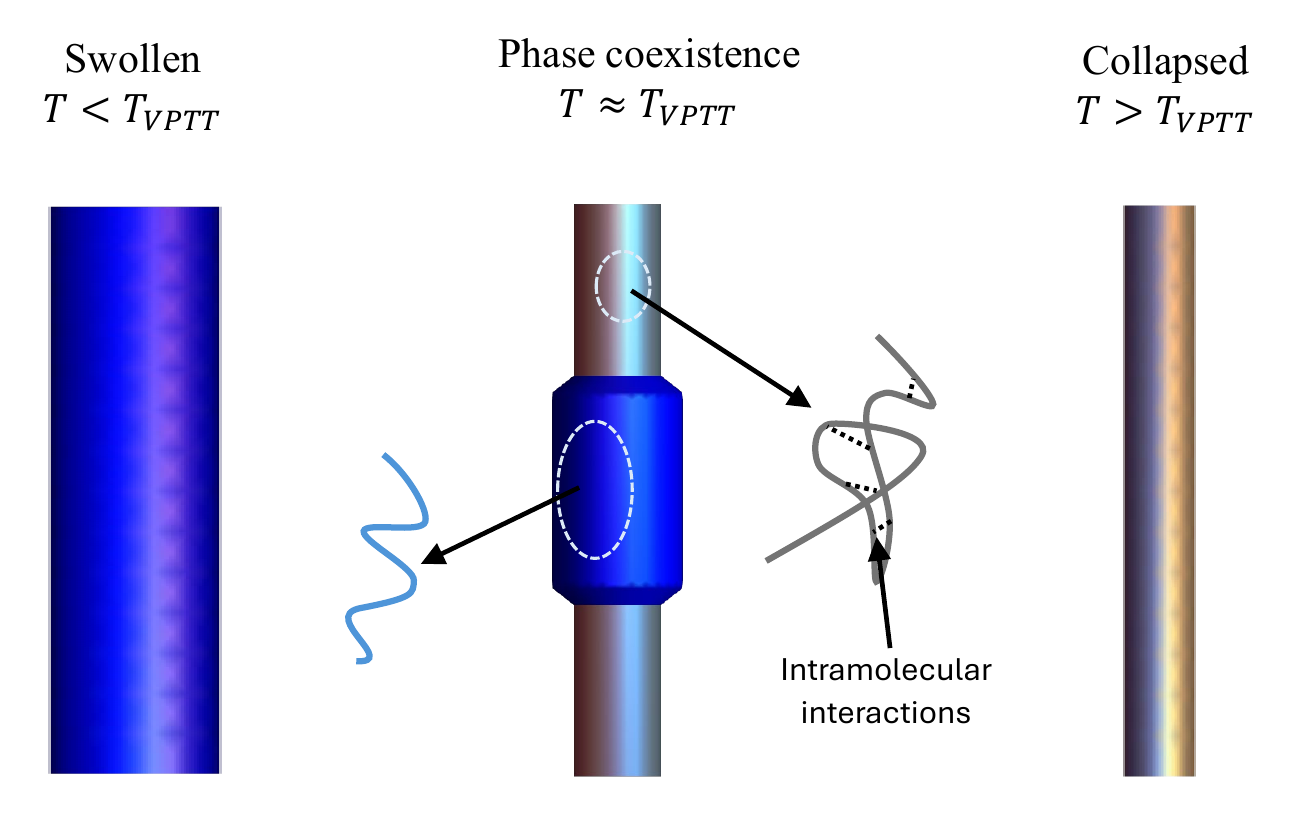}\hspace*{\fill}

\caption{The swollen phase, phase coexistence, and the collapsed phase of a
PNIPAM rod. \label{fig:coexistence_demonstration}}
\end{figure}

We recall that a PNIPAM hydrogel that is subjected to a temperature
$T<\Tvptt$ is in a swollen state while at temperatures $T>\Tvptt$
the network is in a collapsed configuration. Phase coexistence, or
the existence of a swollen and a collapsed phase simultaneously, in
PNIPAM rods was reported to occur in response to a mechanical force
around the VPTT (i.e. $T\approx\Tvptt$) by \citet{Suzuki&ishii99JCP}.
The phase coexistence is illustrated in Fig. \ref{fig:coexistence_demonstration}.
To demonstrate this phenomenon, \citet{Suzuki&ishii99JCP} measured
$\Tvptt\approx33.5^{\circ}C$ and performed two types of experiments
with different boundary conditions at $T\approx\Tvptt$ - (1) time-dependent
phase coexistence in fixed PNIPAM rods and (2) stretch-induced phase
coexistence. 

In the first experiment, a swollen PNIPAM rod was fixed to a constant
(relaxed) length at $T=30^{\circ}C$ and heated up to $T=33.5^{\circ}C$.
At the beginning of the experiment ($t=0$), the rod was fully swollen.
As time progressed, a collapsed phase began to develop in the swollen
rod and slowly grow until a steady state was achieved. After 200 hours,
the rod exhibited a stable configuration with phase coexistence that
lasted several days. In the second experiment, a collapsed PNIPAM
rod was placed at a constant temperature $T=33.5^{\circ}C$ and uniaxially
stretched. Under small stretches, the rod elongated while remaining
in a collapsed configuration. Once a critical stretch was applied,
a swollen domain developed in the rod with a size that further increased
with stretch. The exertion of a sufficiently large force led to the
complete transition of the rod into a single-phase swollen configuration.
It is worth mentioning that in both experiments raising or decreasing
the temperature by $0.1^{\circ}C$ led to the full transition of the
rod from one state to another. 

\begin{figure}
\hspace*{\fill}\includegraphics[width=12cm]{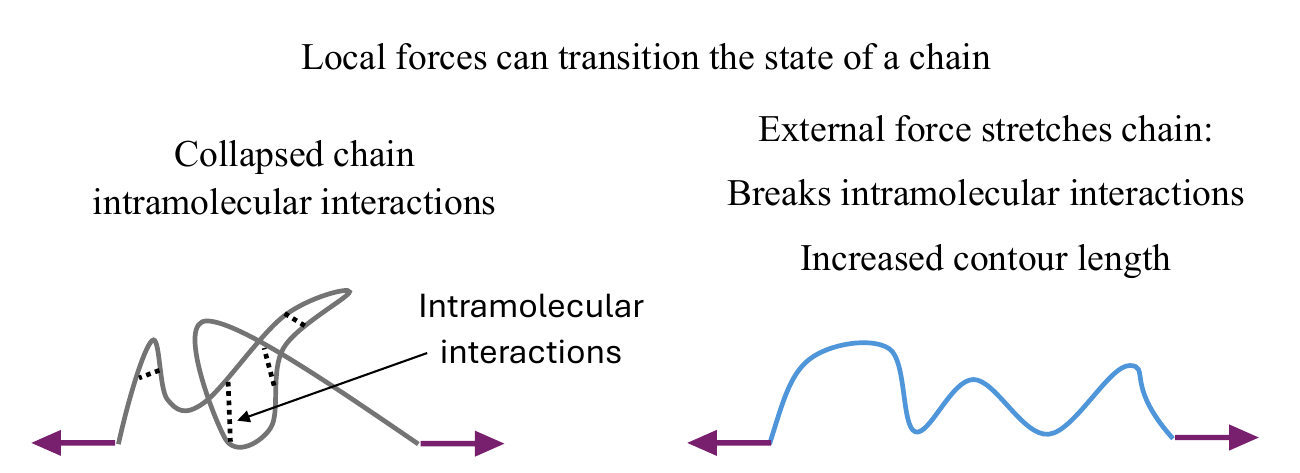}\hspace*{\fill}

\caption{The stretching of a \textquotedblleft collapsed\textquotedblright{}
chain to induce an extended chain by breaking the intramolecular bonds
at temperature $T>\protect\Tvptt$. \label{fig:mechanics_stretch_chain}}
\end{figure}

The origin of this phenomenon is in the mechanically induced local
transition of chains from a globule-like to a coil conformation, depicted
in Fig. \ref{fig:mechanics_stretch_chain}. To understand this, recall
that chains in the collapsed (globule-like) state form intramolecular
interactions that develop between spatially close hydrophilic side
groups and act as physical cross-links \citep{deshmukh&etal13Polymer,Tavagnacco&etal18PCCP,sun&etal04ACIE}.
These interactions essentially cause the chain to fold and shield
the hydrophobic groups from water by assembling hydrophobic clusters
\citep{cho03Macromol,Abbott&etal15JPCB,Levin2025}. As a consequence,
collapsed chains are hydrophobic, have a shorter effective contour
length, and limited mobility. The application of a sufficiently large
local force on the chain can disrupt and break the intramolecular
bonds, exposing the hydrophilic side groups \citep{sun&etal04ACIE,Haupt&etal02lang,Li&walker10JACS,deshmukh&etal13Polymer,Pang&cui13Lang,Liang&etal22Macromol,Levin2025}.
This results in two main consequences - the chain becomes hydrophilic
and attracts water molecules and the contour length of the chain increases. 

At the network level, the mechanically-induced transition of a sufficient
number of chains from one state to another marks the nucleation of
a local transition. For example, stretching a collapsed rod around
the VPTT leads to the local extension of chains, which transition
from globule-like to coil. In turn, the exposure of the hydrophilic
amide side-groups increases attraction to water molecules and motivates
water uptake, ultimately leading to the formation of a swollen domain
in a collapsed PNIPAM rod. In the case of a swollen network that is
fixed in length in an environment with a temperature around the VPTT,
the thermal energy works towards contracting the rod. The thermal
energy works towards transitioning the chains to a collapsed state,
and over time collapsed domains form in the swollen bulk. 

This work aims to provide a better understanding of the underlying
mechanisms that enable phase coexistence in PNIPAM networks. In the
following, I begin by developing a constitutive local model for the
collapsed and the swollen chains and networks. Next, I phrase the
conditions for the coexistence of phases and compare our model predictions
to experimental findings. 

\section{Collapsed versus swollen phases in PNIPAM networks}

The work of \citet{Levin2025} modeled the mechanical response of
PNIPAM networks and the temperature-induced swollen-to-collapsed transition
by introducing explicit relations that capture the dependency of the
effective contour length and the number of repeat units in a chain
on temperature. The aim of the current work is to explain the origin
of the coexistence of phases around the VPTT, and therefore I begin
by developing constitutive relations for the chain and the network
response in each of the phases separately. The chains are modeled
as freely-jointed with contour lengths and number of repeat units
that vary in the coil and the globule-like states. 

First, we recall the governing equations of polymer networks comprising
freely-jointed chains. The entropy of a freely-jointed chain with
$\n$ repeat units and a contour length $\L$ \citep{Kuhn1942,Flory53book,Treloar75book}
\begin{equation}
\entropyC\left(\rat\right)=\mathrm{const}-\boltzmann\n\left(\rat\lang\left(\rat\right)+\ln\left(\frac{\lang\left(\rat\right)}{\sinh\left(\lang\left(\rat\right)\right)}\right)\right),
\end{equation}
where $\boltzmann$ is the Boltzmann constant, $\rat=\etoe/\L$ is
the ratio between the end-to-end distance and the contour length of
the chain, and the function $\lang$ is determined from the Langevin
function $\rat=\coth\lang-1/\lang$, which can be approximated via
$\lang\approx\rat\left(3-\rat^{2}\right)/\left(1-\rat^{2}\right)$
\citep{Cohen91RA}.

The free energy associated with the chains can be written as $\freeC=-T\entropyC$,
and accordingly the force required to maintain an end-to-end distance
$\etoe$ is \citep{Treloar75book,Flory53book} 
\begin{equation}
\forceC\left(\rat,\n\right)=\frac{\partial\freeC}{\partial\etoe}=\boltzmann T\frac{\n}{\L}\lang\left(\rat\right).\label{eq:force_chain}
\end{equation}

To model the macroscopic response, consider a network with $\chains$
chains per unit dry volume subjected to a mechanical force at temperature
$T$. As a result of a thermo-mechanical loading, the network experiences
the macroscopic deformation gradient $\defgradT$. To integrate from
the chain to the network level, consider a chain with an initial end-to-end
vector $\RT=R\RhT$, where $\R=\L/\sqrt{\n}$ is the initial end-to-end
distance and $\RhT$ is the direction. Following common practice \citep{Treloar75book,Flory53book},
I assume that the chains experience the macroscopic deformation gradient
such that the deformed end-to-end vector is $\rT=\defgradT\RT$ with
a length $\r$ and the interactions between the chains are negligible.
Accordingly, the total entropy of a referential volume element $\d V_{0}$
is 
\begin{equation}
\entropy\left(\defgradT,\R\right)=\frac{1}{\d V_{0}}\sum_{i}\entropyC^{\left(i\right)}=\chains\left\langle \entropyC\right\rangle ,
\end{equation}
where the summation is carried over all chains and $\left\langle \entropyC\right\rangle $
is the average entropy. Henceforth, the average of a quantity $\bullet$
is denoted by $\left\langle \bullet\right\rangle =\left(\sum_{i}\entropyC^{\left(i\right)}\right)/\chains\,\d V_{0}$.
Subsequently, the total energy free energy-density of the network
due to the deformation of the chains is 
\begin{equation}
\freen\left(\defgradT,\R,T\right)=-T\entropy.\label{eq:network_energy}
\end{equation}

In the following, we employ the above formulation to develop the energy-density
functions for the swollen ($T<\Tvptt$) and the collapsed ($T>\Tvptt$)
states in PNIPAM networks. 

\subsection{The swollen state ($T<\protect\Tvptt$)}

Consider a dry PNIPAM network under a temperature $T<\Tvptt$ comprising
$\chains$ chains per unit dry volume. In the undeformed dry state,
the material points are denoted by $\refptsT_{s}$. The network is
placed in an aqueous bath and allowed to swell under mechanical forces.
Due to the water uptake and the application of a mechanical load,
the hydrogel deforms such that in the current configuration its material
points are denoted by $\curptsT_{s}$. The deformation gradient is
defined as $\defgradTs=\partial\curptsT_{s}/\partial\refptsT_{s}=\Js^{1/3}\defgradTsIso$,
where $\det\defgradTs=\Js$ is the volumetric deformation due to swelling
and $\det\defgradTsIso=1$ is the deformation due to the isochoric
distortion of the network. The density of water molecules per unit
dry volume in the swollen gel is denoted $\water$. 

From a microscopic viewpoint, the average chain comprises $\nmax$
repeat units with a contour length $\Lmax$, and accordingly the average
end-to-end distance is $\R_{s}=\Lmax/\nmax$. The force on a chain
is given by $\forces=\forceC\left(\ratm,\nmax\right)$, where $\forceC$
is given in Eq. \ref{eq:force_chain} and $\ratm=r/\Lmax$. 

The total energy-density of the swollen network can be written as
\citep{cohen&mcmeeking19JMPS}
\begin{equation}
\free_{s}\left(\defgradTs,T\right)=\freen\left(\defgradTs,\R_{s},T\right)+\freem\left(\Js,T\right)-\prs\left(\Js-\frac{1}{\cp}\right),
\end{equation}
where the first term is given in Eq. \ref{eq:network_energy} with
$\n=\nmax$, the second contribution is the energy of mixing \citep{Huggins1942,Flory1942}
\begin{equation}
\freem=\boltzmann T\left(\nmax\ln\left(1-\cp\right)+\chains\ln\cp+\chi\,\nmax\,\cp\right),
\end{equation}
where $\cp=1/\Js$ is the volume fraction of the polymer in the hydrogel
and $\chi$ is the dimensionless interaction parameter associated
with the heat of mixing that accounts for the solvent-network interactions,
and $\prs$ is a Lagrange multiplier that enforces the incompressibility
of the polymer and the water molecules and accounts for the osmotic
pressure.

The true stress associated that develops in the swollen hydrogel is
\begin{equation}
\stressTs=\frac{1}{\Js}\frac{\partial\free_{s}}{\partial\defgradTs}\defgradTs^{T}=\frac{\chains}{\Js}\left\langle \stressTc^{\left(s\right)}\right\rangle -\prs\mathbf{I},\label{eq:stress_s}
\end{equation}
where 
\begin{equation}
\stressTc^{\left(s\right)}=\forces\frac{\R_{s}^{2}}{\r}\defgradTs\RhT\otimes\defgradTs\RhT,\label{eq:stress_chain_swollen}
\end{equation}
is the stress on a chain. 

The chemical potential of a water molecule in the network is
\begin{equation}
\chempot=\frac{\partial\free_{s}}{\partial\water}=\chempot_{0}+\boltzmann T\left(\ln\left(1-\cp\right)+\cp+\chi\cp^{2}\right)+\prs\volw,\label{eq:chemical_potential_s}
\end{equation}
where $\volw$ is the volume of a water molecule and $\chempot_{0}$
is the reference chemical potential of a water molecule in the aqueous
bath.

\subsection{Collapsed state ($T<\protect\Tvptt$)}

The local behavior of a chain in the collapsed phase is inherently
different than that in the swollen state. Specifically, as shown in
Fig. \ref{fig:mechanics_stretch_chain}, the application of a sufficiently
large force leads to the pulling-out of chain segments from the collapsed
globule, resulting in a random coil conformation \citep{Liang&naka18,Li&walker10JACS,Haupt&etal02lang}.
In addition, the collapsed PNIPAM network is hydrophobic and therefore
swelling does not occur. In the following, the free energy-density
of a collapsed gel that captures these two effects is developed.

Consider a dry PNIPAM network at a temperature $T>\Tvptt$ comprising
$\chains$ chains per unit dry volume. An external force is applied
and the material deforms. The material points in the undeformed and
the deformed configurations are denoted by $\refptsT_{g}$ and $\curptsT_{g}$,
respectively, such that the deformation gradient is $\defgradTg=\partial\curptsT_{g}/\partial\refptsT_{g}$.
The network is assumed to be incompressible, and therefore $\det\defgradTg=1$. 

\begin{figure}
\hspace*{\fill}\includegraphics[width=10cm]{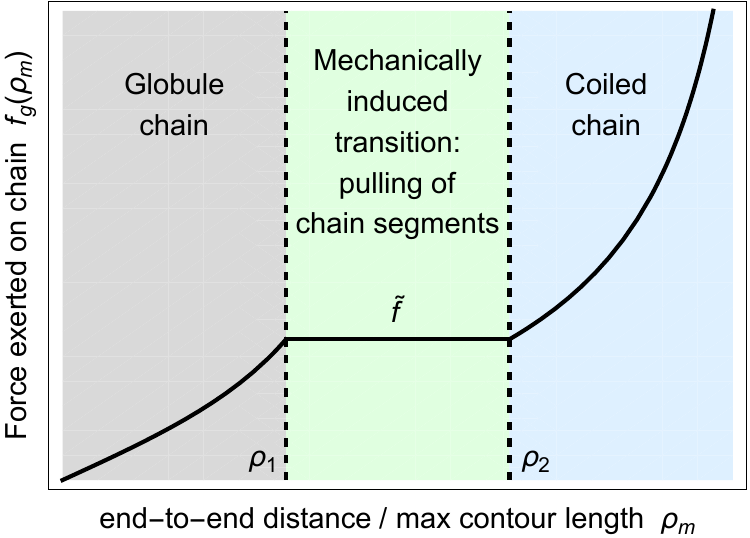}\hspace*{\fill}

\caption{The force $\protect\forceg$ exerted on a collapsed chain as a function
of the ratio $\protect\ratm$. \label{fig:collapsed_chain_force}}
\end{figure}

From a microscopic viewpoint, the chains are initially in a globule
(collapsed) state in which intramolecular bonds inhibit chain mobility.
On average, the chains have $\nmin$ repeat units with an effective
contour length $\Lmin$ such that the end-to-end distance is $\R_{g}=\Lmin/\nmin$.
Once a sufficiently large force $\fcr$ is applied, the intramolecular
bonds break (or, alternatively, chain segments are being pulled out)
to release a ``hidden-length''. As a result, chains gain additional
mobility (or additional degrees of freedom) such that they comprise
$\nmax>\nmin$ repeat units with a contour length of $\Lmax>\Lmin$.
As a result, the end-to-end distance of the chain jumps from $\etoe_{1}$
to $\etoe_{2}$.

To capture this dependency, we define the ratio $\ratm=\etoe/\Lmax$
as the ratio between the end-to-end distance and the contour length
of the fully extended random coil (i.e. the chain in the swollen state),
which enables one to write the ratio $\rat_{g}=\etoe/\Lmin=\ratm\Lmax/\Lmin$.
This allows to express the entropy and the local force on the chain
(Eq. \ref{eq:force_chain}) below and above the critical force $\fcr$
in terms of $\ratm$. Specifically, the force on a chain is given
via the piece-wise function \citep{Olive2024}
\begin{equation}
\forceg\left(\ratm\right)=\boltzmann T\begin{cases}
\frac{\nmin}{\Lmin}\lang\left(\ratm\frac{\Lmax}{\Lmin}\right) & 0\le\ratm<\ratmin\\
\fcr & \ratmin<\ratm<\ratmax\\
\frac{\nmax}{\Lmax}\lang\left(\ratm\right) & \ratmax\le\ratm<1
\end{cases},\label{eq:force_chain_globule}
\end{equation}
where $\ratmin=\etoe_{1}/\Lmax$ and $\ratmax=\etoe_{2}/\Lmax$ are
the ratios between the end-to-end distance and the extended contour
length before and after the dissociation of the intramolecular bonds,
respectively. The force-elongation response of a collapsed chain is
depicted in Fig. \ref{fig:collapsed_chain_force}. 

Consequently, the energy density of the collapsed PNIPAM network is
\begin{equation}
\free_{g}=\freen\left(\defgradTg,\R_{g},\nmax,\nmin,T\right),
\end{equation}
and accordingly the stress is
\begin{equation}
\stressTg=\frac{\partial\free_{g}}{\partial\defgradTg}\defgradTg^{T}=\chains\left\langle \stressTc^{\left(g\right)}\right\rangle -\prg\mathbf{I},\label{eq:stress_g}
\end{equation}
where 
\begin{equation}
\stressTc^{\left(g\right)}=\forceg\frac{\R_{g}^{2}}{\r}\defgradTg\RhT\otimes\defgradTg\RhT,\label{eq:stress_chain_collapsed}
\end{equation}
is the stress on a collapsed chain and $\prg$ is a pressure-like
term that enforces the incompressibility of the network.

\section{Mechanisms behind the coexistence of phases}

\global\long\def\reflen{B}%
\global\long\def\refleng{B_{g}}%
\global\long\def\reflens{B_{s}}%
\global\long\def\areag{A_{g}}%
\global\long\def\areas{A_{s}}%
\global\long\def\stretchmin{\stretch_{1}}%
\global\long\def\stretchmax{\stretch_{2}}%

\global\long\def\cdf{P}%
\global\long\def\ratd{\rho_{d}}%
\global\long\def\dev{v}%
\global\long\def\ang{\theta}%
\global\long\def\azi{\varphi}%
\global\long\def\marginal{q}%
\global\long\def\pang{p_{\ang}}%
\global\long\def\openc{N_{e}}%

To better understand the underlying mechanisms that enable phase coexistence
in PNIPAM gels, we follow the work of \citet{Suzuki&ishii99JCP} and
examine the uniaxial stretching of a collapsed rod around the VPTT.
In the collapsed network, local extension of chains motivates the
globule to coil transition by breaking intramolecular bonds and ``pulling''
chain segments out of the globule, thereby inducing elongated hydrophilic
chains that attract water \citep{Liang&naka18,Li&walker10JACS,Haupt&etal02lang}.
Once a sufficient number of chains ``open up'' in the vicinity of
one another, a local swollen domain emerges. This process is stochastic
in nature, with a probability of extended chains that increases with
the external force. Accordingly, swollen domains can form in a collapsed
network, resulting in phase coexistence within the PNIPAM rod.

In this section, a model for the formation of phase coexistence is
derived based on the above description.

\subsection{Kinematics}

\begin{figure}
\hspace*{\fill}\includegraphics[width=12cm]{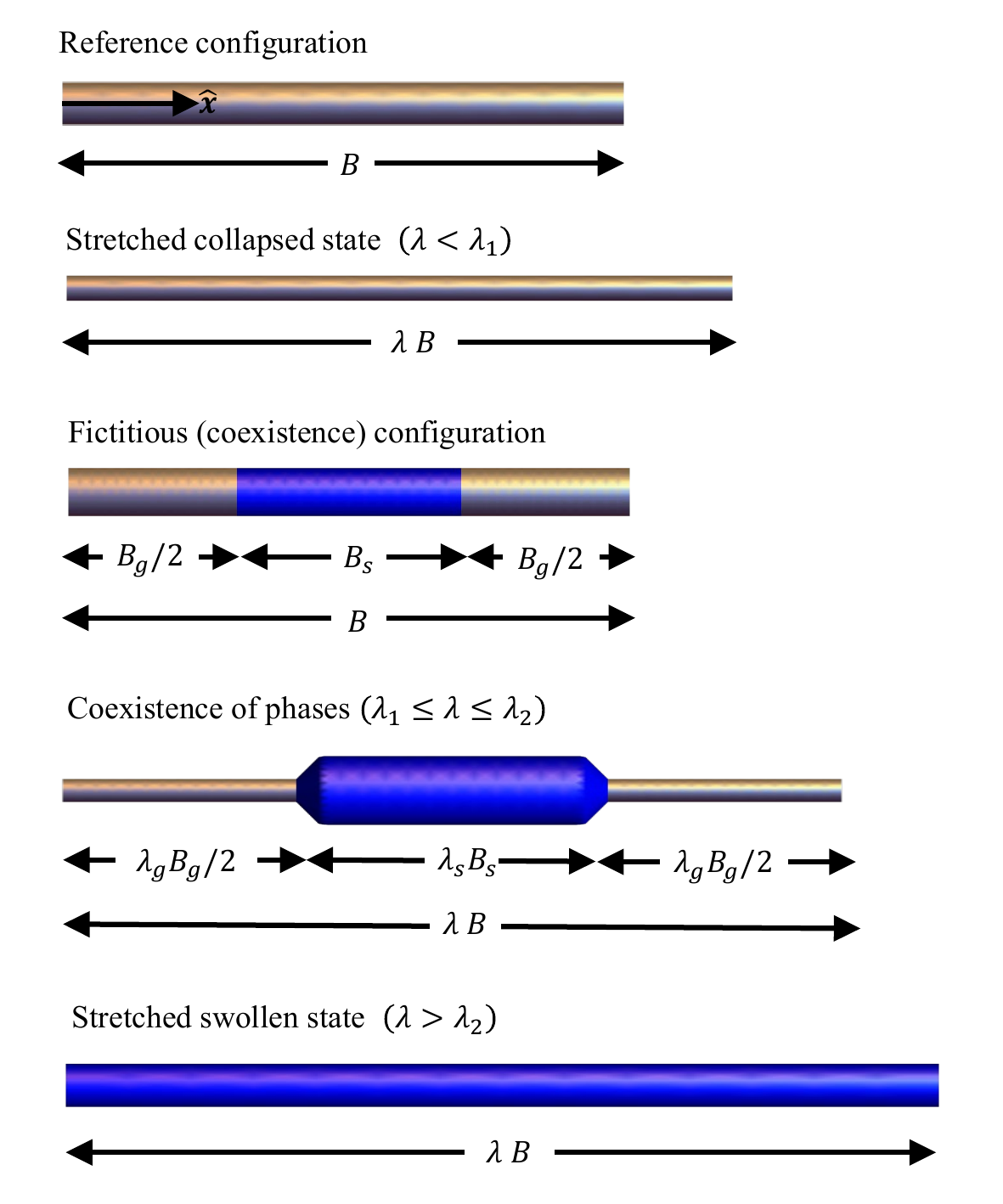}\hspace*{\fill}

\caption{A schematic illustration of the stretching of a collapsed PNIPAM chain:
a collapsed chain is stretched from the reference configuration by
$\protect\stretch<\protect\stretchmin$. Further extension ($\protect\stretchmin<\protect\stretch<\protect\stretchmax$)
leads to the coexistence of phases - the rod is stretched from a fictitious
to a deformed state comprising collapsed and swollen domains. At $\protect\stretch=\protect\stretchmax$
the rod transitions completely to the swollen state, and further stretch
($\protect\stretch>\protect\stretchmax$) leads to the elongation
of a swollen rod. \label{fig:coexistence_diagram}}
\end{figure}

Following the experiment of \citet{Suzuki&ishii99JCP}, consider a
collapsed PNIPAM rod of length $\reflen$ and cross-sectional area
$\areag$. The network comprises $\chains$ chains per unit dry volume
(i.e. above the VPTT). We define a coordinate system $\left\{ \xh,\yh,\zh\right\} $
in which the rod is aligned along the $\xh$-direction. The rod stretches
such that its overall length is $\stretch\reflen$, where $\stretch$
is the axial stretch. 

To capture phase coexistence, we define the critical stretches $\stretchmin$
and $\stretchmax$, such that for stretches $\stretch<\stretchmin$
the rod is in a fully collapsed state and stretches $\stretch>\stretchmax$
result in a fully swollen rod. In the range $\stretchmin\le\stretch\le\stretchmax$,
phase coexistence is observed. The origin of this behavior can be
understood from the microstructure - for longitudinal stretches $\stretch<\stretchmin$,
the local forces are not sufficient to transition enough chains from
a globule-like to a coiled conformation in order to trigger a phase
transition. Once a stretch $\stretch=\stretchmin$ is prescribed,
many chains ``open up'' and become hydrophilic, leading to significant
water uptake and the initiation of a swollen domain. Further increase
in stretch (or external force) motivates the local transition of additional
chains, thereby increasing the size of the swollen domains in the
rod. At $\stretch=\stretchmax$, all chains transitioned to coils
and the rod is fully swollen. This process is depicted in Fig. \ref{fig:coexistence_diagram}.

For any stretch $\stretchmin\le\stretch\le\stretchmax$, it is convenient
to define a fictitious unloaded state in which the length of the collapsed
and the swollen sections of the rod are $\refleng=\refleng\left(\stretch\right)$
and $\reflens=\reflens\left(\stretch\right)$, respectively, such
that $\reflen=\refleng+\reflens$. For completeness, it is stated
that $\refleng\left(\stretch<\stretchmin\right)=\reflen$ and $\reflens\left(\stretch<\stretchmin\right)=0$
(i.e. the rod is in a collapsed state for $\stretch<\stretchmin$)
and $\refleng\left(\stretch>\stretchmax\right)=0$ and $\reflens\left(\stretch>\stretchmax\right)=\reflen$
(i.e. the rod is in a swollen state for $\stretch>\stretchmax$).
Next, the rod is stretched from this fictitious state to a length
$\stretch\reflen=\stretchg\refleng+\stretchs\reflens$, where $\stretchg$
and $\stretchs$ are the local stretches of the collapsed and the
swollen regions, respectively (see Fig. \ref{fig:coexistence_diagram}). 

Since the rod is uniaxially stretched, the deformation gradient of
the the collapsed and the swollen domains can be written as 
\begin{equation}
\defgradTg=\stretchg\xh\otimes\xh+\frac{1}{\sqrt{\stretchg}}\left(\yh\otimes\yh+\zh\otimes\zh\right),
\end{equation}
and

\begin{equation}
\defgradTs=\Js^{1/3}\left(\stretchs\xh\otimes\xh+\frac{1}{\sqrt{\stretchs}}\left(\yh\otimes\yh+\zh\otimes\zh\right)\right),
\end{equation}
respectively. 

\subsection{Microstructural conditions for formation of phases }

To model the nucleation, or the initial formation of a swollen phase
in a collapsed rod, recall that a force $\fcr$ must be applied locally
on a chain in order to ``pull-out'' chain segments and break the
globule structure \citep{Liang&naka18,Li&walker10JACS,Haupt&etal02lang}.
Following Eq. \ref{eq:force_chain_globule}, this force corresponds
to a ratio $\ratmin\le\ratm\le\ratmax$ between the end-to-end distance
and $\Lmax$, and therefore we expect the transition of a chain to
occur in that range. To form a swollen domain in the network, a sufficient
number of chains must ``open up'' to allow for water uptake. 

\begin{figure}
\hspace*{\fill}\includegraphics[width=10cm]{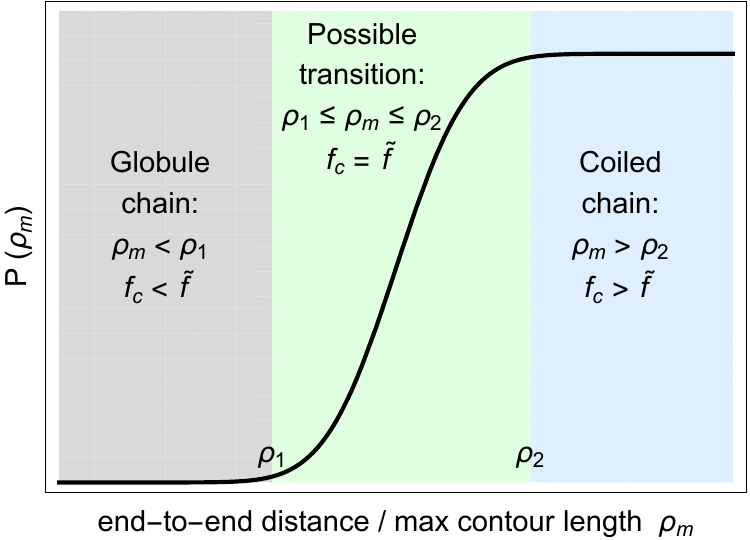}\hspace*{\fill}

\caption{Illustration of the probability $\protect\cdf$ as a function of the
ratio $\protect\ratm$ that a chain transitions from globule to coil.
\label{fig:chain_transition_probability}}
\end{figure}

To capture the stochastic nature of the globule-to-coil transition,
an appropriate probability of transition $0\le\cdf\le1$ must be defined
for which $\cdf\left(\ratm<\ratmin\right)=0$, corresponding to a
collapsed chain, and $\cdf\left(\ratm>\ratmax\right)=1$, corresponding
to a chain in an extended state. In this work, I employ a continuous
truncated Gaussian-distribution based cumulative distribution function
\begin{equation}
\cdf\left(\ratm\right)=\frac{\mathrm{Erf}\left(\frac{\ratm-\ratd}{\sqrt{2}\dev}\right)+\mathrm{Erf}\left(\frac{\ratd}{\sqrt{2}\dev}\right)}{\mathrm{Erf}\left(\frac{1-\ratd}{\sqrt{2}\dev}\right)+\mathrm{Erf}\left(\frac{\ratd}{\sqrt{2}\dev}\right)},\label{eq:probability_single_chain}
\end{equation}
where $0<\ratd<1$ is a parameter that governs the ratio at which
the chain transitions and $\dev$ is the standard deviation. The probability
is illustrated in Fig. \ref{fig:chain_transition_probability}. 

The nucleation of a swollen domain requires the ``pulling out''
of chains in the network. One can compute the fraction of open chains
as follows: first, we denote by $\ratm^{\left(i\right)}\left(\ang^{\left(i\right)}\right)$
chains with a referential end-to-end direction $\RhT^{\left(i\right)}$
that form an angle $\ang^{\left(i\right)}\le\ang<\ang^{\left(i\right)}+\Delta\theta$
with the longitudinal direction $\xh$. The fraction of these chains
in the network is proportional to the solid angle such that $\pang\left(\ang\right)=\sin\left(\ang\right)\Delta\ang/2$,
where $0\le\ang\le\pi$. As the rod stretches, chains deform and one
can write $\ratm^{\left(i\right)}=\ratm^{\left(i\right)}\left(\ang^{\left(i\right)},\defgradT\right)$.
Note that chains with $\ang^{\left(i\right)}\rightarrow0$ and $\ang^{\left(i\right)}\rightarrow\pi$
extend, while chains along the transverse direction (i.e. $\ang^{\left(i\right)}\approx\pi/2$)
shorten, with deformations that are proportional to the local deformation
gradient. As expected, the stretching of the chains increases the
likelihood of ``opening up'' to an extended conformation (see Eq.
\ref{eq:probability_single_chain}). The density of extended chains
per unit dry volume along the $i$-th direction is given by $\openc^{\left(i\right)}=\cdf\left(\ratm^{\left(i\right)}\right)\pang\left(\ang\right)\chains$.
Subsequently, one can compute the fraction of open chains in the network
via 
\begin{equation}
0\le\marginal=\frac{\sum_{i}\openc^{\left(i\right)}}{\chains}=\int_{0}^{\pi}\cdf\left(\ratm^{\left(i\right)}\right)\pang\left(\ang\right)\d\ang\le1,\label{eq:probability}
\end{equation}
where in passing the summation is converted to integration over all
angles $0\le\ang\le\pi$. 

Before proceeding, it is worthwhile noting that chains along the plane
transverse to the stretching direction shorten and therefore cannot
naturally open up. The transition of these chains is enabled by the
following mechanism: as chains get pulled out to an extended conformation,
they attract water molecules from the environment, resulting in local
swelling. The local presence of water molecules motivates all chains
in the network to transition, ultimately leading to a macroscopically
visible swollen phase. 

To relate the local response to the macroscopic behavior, we conjecture
that the fraction per length of the swollen domain is proportional
to $\marginal$ such that $\marginal\approx\reflens/\reflen$. By
employing the relation $\stretch\reflen=\stretchg\refleng+\stretchs\reflens$,
we relate the macroscopic deformation $\stretch$ to the local deformations
$\stretchs$ and $\stretchg$ of the swollen and collapsed domains
via
\begin{equation}
\stretch=\stretchg\left(1-\marginal\right)+\stretchs\marginal.\label{eq:kinematic_stretch_relation}
\end{equation}
Note that in the limit $\marginal=0$ the stretch $\stretch=\stretchg$,
corresponding to a collapsed rod, whereas in the limit $\marginal=1$
the network is in its swollen configuration and the stretch $\stretch=\stretchs$. 

\subsection{Equilibrium conditions\label{subsec:Equilibrium-conditions}}

A rod in phase coexistence must satisfy mechanical and chemical equilibrium.
Mechanical equilibrium requires that the stress in the two phases
be divergence free, i.e. $\nabla\cdot\stressTs=\mathbf{0}$ and $\nabla\cdot\stressTg=\mathbf{0}$.
In addition, the longitudinal forces along the rod in the collapsed
and the swollen domains must be equal, i.e. $\int\stressTg\xh\cdot\xh\d a_{g}=\int\stressTs\xh\cdot\xh\d a_{s}$.
The true stress in the swollen domains $\stressTs$ and the collapsed
domains $\stressTg$ is given in Eqs. \ref{eq:stress_s} and \ref{eq:stress_g},
respectively. To determine the deformed areas $a_{g}$ and $a_{s}$
in the collapsed and the swollen states, we assume that the two phases
are incompressible and employ Nanson's formula. In the case of uniaxial
extension, the deformed areas are $a_{g}=A_{g}/\stretchg$ and $a_{s}=A_{s}/\stretchs$,
where $A_{g}$ and $A_{s}$ are the areas of the fully collapsed and
the fully swollen rods in the traction free configuration. These areas
were experimentally measured by \citet{Suzuki&ishii99JCP} and \citet{Suzuki&Sanda97JCP}. 

Due to the hydrophilicity, the swollen domain must also satisfy chemical
equilibrium, which pertains to the interactions between the solvent
molecules and the gels. This is satisfied via $\chempot=\chempot_{0}$
(see Eq. \ref{eq:chemical_potential_s}). 

\subsection{Integration from the chain to the network level}

To integrate from the chain to the network level, I employ the numerical
micro-sphere technique \citep{Bazant&oh86ZAMM,Miehe2004,Levin2024,cohen&mcmeeking19JMPS}.
This method enables one to estimate the directional averaging of one-dimensional
elements (the end-to-end vector directions $\RhT$) over a unit sphere
to obtain macroscopic quantities. Specifically, the stress that develops
in a unit volume $\d V_{p}$ with $\chains$ chains per unit volume
can be determined via 
\begin{equation}
\left\langle \stressTc\right\rangle =\frac{1}{4\pi}\int_{A}\stressTc\d A\approx\sum_{i=1}^{m}\stressTc^{\left(i\right)}w^{\left(i\right)},
\end{equation}
where the summation is carried over $m$ representative directions
and the index $i=1,2,...,m$ refers to the $i$-th chain with an initial
end-to-end direction $\RhT^{\left(i\right)}$ and a non-negative weight
$w^{\left(i\right)}$. Here, $\stressTc=\stressTc^{\left(s\right)}$
(Eq. \ref{eq:stress_chain_swollen}) or $\stressTc=\stressTc^{\left(g\right)}$
(Eq. \ref{eq:stress_chain_collapsed}) to account for the stress of
chains in the swollen or the collapsed domains, respectively. 

It is emphasized that the weights are constrained such that $\sum_{i}w^{\left(i\right)}=1$.
In addition, in isotropic networks such as the PNIPAM hydrogels, the
representative directions must be chosen such that $\sum_{i}\RhT^{\left(i\right)}w^{\left(i\right)}=\mathbf{0}$
and $\sum_{i}\RhT^{\left(i\right)}\otimes\RhT^{\left(i\right)}w^{\left(i\right)}=1/3\mathbf{I}$.
\citet{Bazant&oh86ZAMM} showed that $m=42$ specific representative
directions and weights (given in Table 1 of that work) provide sufficient
accuracy for isotropic materials, and this conclusion is used in this
work.

\section{Comparison to experimental findings}

\global\long\def\ds{d_{s}}%
\global\long\def\dg{d_{g}}%
\global\long\def\shear{G}%
\global\long\def\As{A_{s}}%
\global\long\def\Ag{A_{g}}%

\begin{table}
\caption{Summary of model parameters \label{tab:model_parameters}}

\centering{}%
\begin{tabular}{cc}
\toprule 
Notation & Definition\tabularnewline
\midrule
\midrule 
$\Lmin/\Lmax$ & Contour length of collapsed / extended chain\tabularnewline
\midrule 
$\nmin/\nmax$ & Number of repeat units in collapsed / extended chain\tabularnewline
\midrule 
$\fcr$ & Critical force to break intramolecular bonds in collapsed chain \tabularnewline
\midrule 
$\ratm$ & Ratio between end-to-end distance $r$ and $\Lmax$\tabularnewline
\midrule 
$\ratmin/\ratmax$ & \begin{cellvarwidth}[t]
\centering
Ratio between end-to-end distance and $\Lmax$ before / after\\
dissociation of intramolecular bonds
\end{cellvarwidth}\tabularnewline
\midrule 
$\Js$ & Volumetric deformation from dry to swollen networks\tabularnewline
\midrule 
$\chains$ & Chain-density per unit dry volume\tabularnewline
\midrule 
$\shear=\chains\boltzmann T$ & Shear modulus of dry network\tabularnewline
\midrule 
$\stretchg/\stretchs$ & Uniaxial stretch in collapsed / swollen network\tabularnewline
\midrule 
$\stressTc/\stressTs$ & True (Cauchy) stress in collapsed / swollen networks\tabularnewline
\midrule 
$\cdf$ & Probability of conformational transition of chain\tabularnewline
\midrule 
$\dev/\ratd$ & Probability related parameters, govern transition\tabularnewline
\midrule 
$\marginal$ & Fraction of open chains in the network\tabularnewline
\bottomrule
\end{tabular}
\end{table}

To validate the model, we summarize all of the model parameters in
Table \ref{tab:model_parameters} for convenience and compare its
predictions to the experimental findings reported in \citet{Suzuki&ishii99JCP}.
To this end, we follow the experimental work of \citet{Liang&naka18}
and the corresponding model of \citet{Levin2025} and set the effective
contour length of the extended and the collapsed chains $\Lmax=100\,\mathrm{nm}$
and $\Lmin=35\,\mathrm{nm}$, respectively. To satisfy the continuity
condition of the force on a collapsed chain (Eq.\ref{eq:force_chain_globule}),
we set the critical force $\fcr=15\,\mathrm{pN}$ \citep{Liang&naka18},
$\ratmin=0.2$, and $\ratmax=0.5$ such that the number of Kuhn segments
in the two states are $\nmax=193$ and $\nmin=55$. 

From a macroscopic viewpoint, \citet{Suzuki&ishii99JCP} measured
the diameters of the cylinder immediately before and after the VPTT
(i.e. in the swollen and the collapsed phases) $\ds=110\,\mathrm{\mu m}$
and $\dg=58\,\mathrm{\mu m}$, respectively. Accordingly, we can calculate
the cross-sectional areas $\As$ and $\Ag$. In addition, the volume
of a water molecule is $\volw=3\cdot10^{-29}\,\mathrm{m^{3}}$. To
estimate the interaction parameter $\chi$, we consider the traction
free swelling of a PNIPAM rod from the collapsed to the swollen configuration.
The volumetric deformation around the VPTT is assumed to be $\Js=\left(\ds/\dg\right)^{3}\approx6.82$,
which enables one to determine $\chi\approx0.552$ by setting $\stressTs=\mathbf{0}$
and $\chempot=\chempot_{0}$ in Eqs. \ref{eq:stress_s} and \ref{eq:chemical_potential_s}.
This result corresponds to previous experimental findings on PNIPAM
rods \citep{shen&etal19SM,Nigro2017}. It is emphasized that this
is an approximation, since the classical swelling theory does not
account for the curtailment of the chain from an extended to a globule
state. 

The stiffness of the collapsed gel $\shear=\chains\boltzmann\Tvptt=22.7\,\mathrm{kPa}$
at the VPTT is fitted to the experimental findings. This value is
within the range of measured stiffness values for PNIPAM, reported
as $\sim1-10\,\mathrm{kPa}$ and $\sim10^{2}-10^{3}\,\mathrm{kPa}$
below and above the VPTT, respectively \citep{takigawa&etal98PGN,haq&etal17MSEC,Lehmann&etal19CPS}.
The stiffness of the swollen network can be approximated via $\shear/\Js^{1/3}$
\citep{cohen&mcmeeking19JMPS,Flory53book,Flory1942}. 

To capture the transition of local domains from collapsed to swollen
(or swollen to collapsed), the standard deviation $\dev=5\cdot10^{-3}$
and the parameter $\ratd=0.115$, pertaining to the transition probability
in Eq. \ref{eq:probability_single_chain}, are obtained from a fit
to the experimental findings. 

In the following, the two experiments performed by \citet{Suzuki&ishii99JCP}
are considered and investigated. 

\subsection{Stretch induced phase coexistence}

We begin by comparing the model predictions to the stretch-induced
phase coexistence demonstrated by \citet{Suzuki&ishii99JCP}. As illustrated
in Fig. \ref{fig:coexistence_diagram}, a collapsed PNIPAM rod was
stretched at a temperature $T\approx\Tvptt$ ($\sim33.5^{\circ}C$).
Once a sufficient critical force was applied, a swollen phase began
to appear. Further increase in force led to a higher ratio of swollen
portion to total length until ultimately the rod transitions to a
completely swollen continuum. It is again pointed out that since the
rod is subjected to uniaxial tension, chains aligned along the longitudinal
direction experience tension and are pulled-out first. 

\begin{figure}
\hspace*{\fill}(a) \includegraphics[width=6cm]{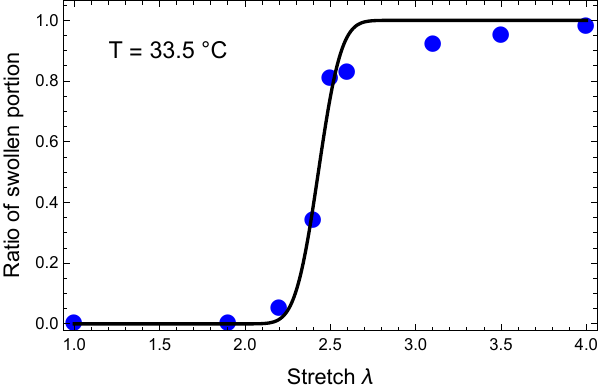}~~~(b)
\includegraphics[width=6cm]{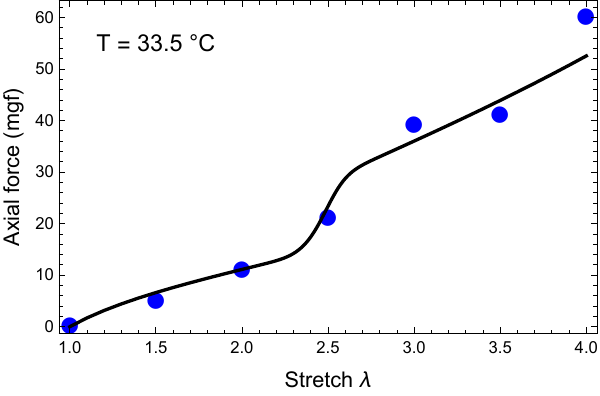}\hspace*{\fill}

\caption{(a) The ratio of the swollen portion and (b) the axial force (in mgf)
as a function of the macroscopic stretch $\protect\stretch$. The
curves correspond to the model predictions and the circular marks
denote the experimental findings reported by (a) \citet{Suzuki&ishii99JCP}
and (b) \citet{Suzuki&Sanda97JCP}. \label{fig:compare_experiment}}
\end{figure}

Fig. \ref{fig:compare_experiment}a plots the ratio of the swollen
portion as a function of the macroscopic stretch $\stretch$ according
to the experimental findings of \citet{Suzuki&ishii99JCP} (circle
marks) and the model predictions (continuous curve). The model agrees
with the experimental data. As shown in the work of \citet{Suzuki&ishii99JCP},
a macroscopic stretch $\stretch\sim2.2$ is required to initiate a
swollen domain and at a stretch $\stretch\sim3.3$ the rod is fully
swollen. 

To further validate the model, we compute the axial force $f=\int\stressTs\xh\cdot\xh\d a_{c}=\int\stressTg\xh\cdot\xh\d a_{s}$
as a function of the stretch $\stretch$. While \citet{Suzuki&ishii99JCP}
did not report the axial force during this experiment, a previous
work (\citet{Suzuki&Sanda97JCP}, by the same author) measured the
force on PNIPAM rods that were held fixed at a stretch $\stretch$
and subjected to increasing temperature. As an estimate, one can take
the forces measured at the temperature $T=33.5^{\circ}\mathrm{C}$
under different stretch values and plot them (circle marks) against
the model predictions (continuous curve), as shown in Fig. \ref{fig:compare_experiment}b.
Once again, the model agrees well with the experimental findings. 

\begin{figure}
\hspace*{\fill}\includegraphics[width=6cm]{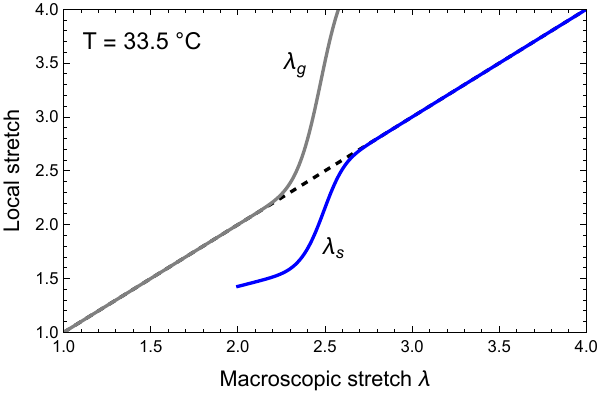}\hspace*{\fill}

\caption{The local stretch in the collapsed ($\protect\stretchg$) and the
swollen ($\protect\stretchs$) domains as a function of the macroscopic
stretch $\protect\stretch$ (dashed line). \label{fig:stretch_distribution}}
\end{figure}

Further insights from the model can be gained by examining the local
stretch in the collapsed ($\stretchg$) and the swollen ($\stretchs$)
domains as a function of the macroscopic stretch $\stretch$, as depicted
in Fig. \ref{fig:stretch_distribution}. Prior to the nucleation of
the swollen phase, the stretch of the collapsed domain is the same
as the macroscopic stretch, i.e. $\stretchg=\stretch$. Once a swollen
domain forms, the chains locally stretch due to swelling and the external
force. We point out that Fig. \ref{fig:stretch_distribution} only
shows the stretch $\stretchs$ associated with a mechanical force,
without considering the swelling-induced extension. 

Interestingly, we find that while the force in the two domains is
equal (to satisfy mechanical equilibrium), the stretch and the stress
differ. To understand this, one must examine the cross-sectional area
and the stiffness of the two domains. With respect to the former,
the cross-sectional area of the swollen regions is larger than that
of the collapsed ones ($\As>\Ag$), as one would expect. Regarding
the latter, the collapsed domains are dry (since the chains transition
to their hydrophobic state) and are therefore stiff. On the other
hand, the stiffness of the swollen domain is influenced by two factors
- the softening due to water uptake and the stretch-induced stiffening
of the chains, which must extend to accommodate water molecules \citep{Flory1942,Flory53book,cohen&mcmeeking19JMPS}.
It is also underscored that the transition of the chain from a collapsed
conformation to an extended coil is accompanied by an increase in
the effective contour length, stemming from the dissociation of the
intramolecular bonds. To ensure equilibrium during phase coexistence,
the stress experienced by the collapsed domains is higher than the
swollen regions, but due to its smaller cross-sectional area we find
that the stretch $\stretchg$ is higher. Once all the chains transition
to an extended state, the rod is in a fully swollen configuration
and $\stretch=\stretchs$. 

\subsection{Relaxation-induced phase coexistence}

\begin{figure}
\hspace*{\fill}\includegraphics[width=12cm]{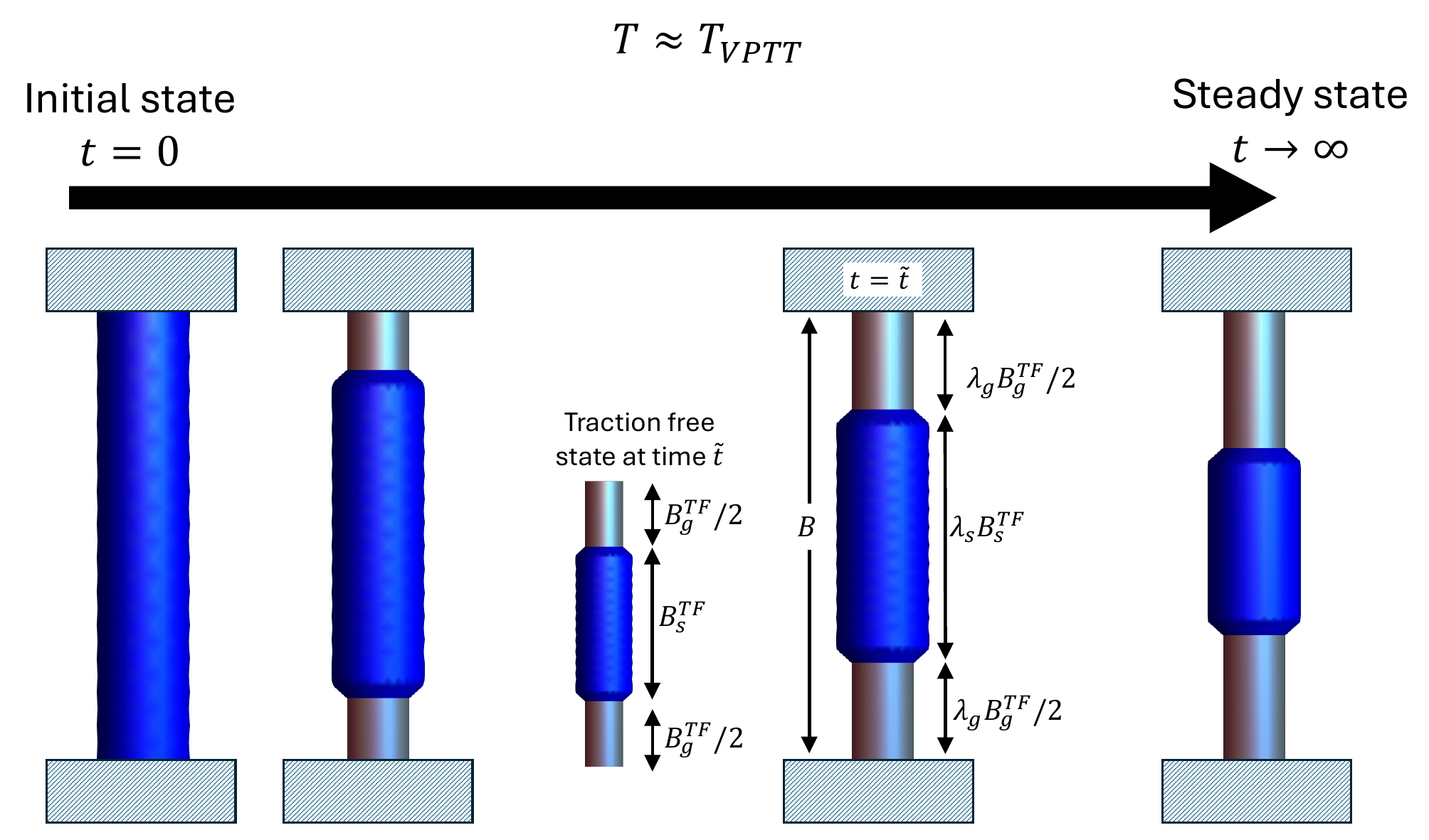}\hspace*{\fill}

\caption{Time dependent coexistence of phases in a constrained PNIPAM rod at
$T\approx\protect\Tvptt$, as shown by \citet{Suzuki&ishii99JCP}.
\label{fig:time_dependent_process}}
\end{figure}

\citet{Suzuki&Sanda97JCP} demonstrated that a swollen PNIPAM rod
held at a fixed length and a critical temperature $T\approx\Tvptt$
($\sim33.5^{\circ}C$) occupies a configuration with phase coexistence
as a function of time (see Fig. \ref{fig:time_dependent_process}).
In this experiment, a swollen rod was held fixed at a constant length
at $T=30^{\circ}C$ and heated to $T=33.5^{\circ}C$. At this point,
the thermal energy motivates the contraction of the chains to achieve
a collapsed rod. In parallel, the external constraint gives rise to
a tensile force that prevents the chains from contracting. As a result,
a rod with phase coexistence develops. It is worth pointing out that
further increase in temperature to $T=33.6^{\circ}C$ led to the disappearance
of phase coexistence and the collapse of the tube. 

As opposed to the previous case, in which a collapsed PNIPAM rod was
stretched to achieve phase coexistence, the rod was initially in the
swollen phase with a fixed length $\reflen$ and collapsed domains
form with time. At the beginning of the experiment, the chains in
the swollen hydrogel have an average end-to-end distance $\R=\Js^{TF}/\sqrt{\nmax}$,
where $\Js^{TF}\approx6.7$ is the volumetric deformation determined
from a traction free rod that is placed in an aqueous bath. Due to
the thermal energy that is transferred to the PNIPAM rod from the
environment, chains begin to collapse an in turn, tensile forces arise.
To understand the time dependent microstructural evolution of the
chains, it is important to note that chains which are aligned along
the fiber direction experience tension due to the tensile force which
prevents them from contracting. On the other hand, chains along the
transverse plane shorten and are significantly more likely to transition
to globule-like conformations. 

Similar to the analysis of a stretch-induced phase coexistence, we
determine the stretches $\stretchs$ and $\stretchg$ of the swollen
and the collapsed domains, respectively, with respect to a fictitious
traction free state. In this configuration, the lengths of the swollen
and the collapsed domains are $B_{s}^{TF}$ and $B_{g}^{TF}$, respectively,
as shown in Fig. \ref{fig:time_dependent_process}. In the beginning
of the experiment ($t=0$), $B_{s}^{TF}+B_{g}^{TF}=B$ and the rod
is traction free. At $t>0$, the length of the traction free rod decreases
$B_{s}^{TF}+B_{g}^{TF}<B$ due to the outward flux of water molecules.
To capture the time-dependent relaxation of the rod, we define the
relaxation stretch from the referential to the traction free configuration
at time $t$, 
\begin{equation}
\stretchr\left(t\right)=\left(1-\stretchr^{\infty}\right)\exp\left(-\frac{t}{\timec}\right)+\stretchr^{\infty},\label{eq:stretch_g_time}
\end{equation}
where $\stretchr\left(t\rightarrow\infty\right)=\stretchr^{\infty}$
is the steady state traction free state and $\timec$ is a characteristic
relaxation time. This constitutive relation can be viewed as rate
of transition of the chains from the swollen to the collapsed phases.
In the following, we fit $\stretchr^{\infty}=0.825$ and $\timec=40\,\mathrm{hrs}$.
The stretch of the collapsed region is defined as the initial and
the current relaxed lengths of the rod, i.e. $\stretchs=1/\stretchr$. 

\begin{figure}
\hspace*{\fill}(a) \includegraphics[width=4.5cm]{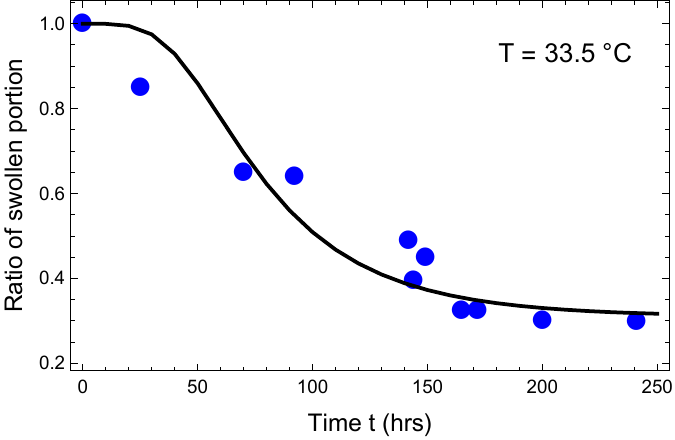}~(b)
\includegraphics[width=4.5cm]{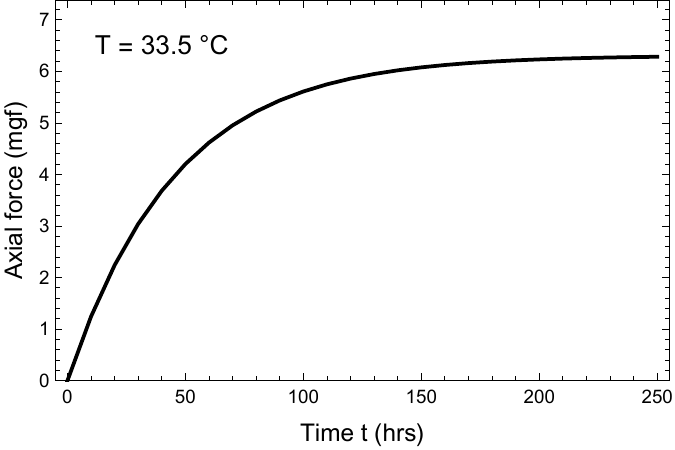}~ (c) \includegraphics[width=4.5cm]{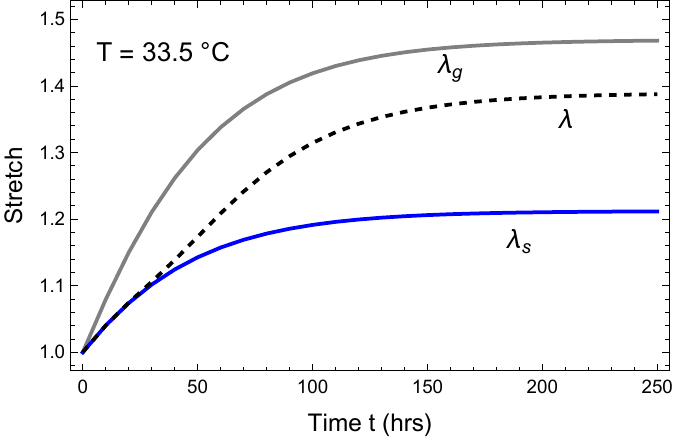}\hspace*{\fill}

\caption{(a) The ratio of the swollen portion, (b) the axial force (in mgf),
and (c) the total stretch $\protect\stretch$ and the local stretch
in the collapsed ($\protect\stretchg$) and the swollen ($\protect\stretchs$)
domains as a function of time $t$ (in hours). The curves correspond
to the model predictions and the circular marks in (a) denote the
experimental data from \citet{Suzuki&ishii99JCP}. \label{fig:fixed_experiment}}
\end{figure}

Fig. \ref{fig:fixed_experiment}a plots the ratio of the swollen portion
as a function of time $t$ (in hours). The circle marks denote the
experimental data from \citet{Suzuki&ishii99JCP} and the continuous
curve is the model prediction. The model is capable of capturing the
experimental results. Experiments revealed that the swollen portion
saturated after $200\,\mathrm{hrs}$ and remained in stable phase
coexistence for several days, indicating a steady state. 

While the tensile force that developed during the experiment was not
reported, it was estimated with the model and is shown in Fig. \ref{fig:fixed_experiment}b.
As expected, the force increases but remains small, corresponding
to the measurements reported by \citet{Suzuki&Sanda97JCP} (see the
circular marks denoted by 1 in Fig. 3a of that work). This prediction
can be validated by comparing the maximum force at steady state $f_{ss}\approx6.3$
based on the model at $t>250\,\mathrm{hrs}$ to that which would be
required to stretch a collapsed rod of length $\reflen_{c}$ to the
original fixed length during the experiment $\reflen$. Here, $\reflen_{c}$
is approximated via the configuration of the swollen gel in the dry
state. Since the rod in the experiment of \citet{Suzuki&ishii99JCP}
achieves stable phase coexistence, the axial force $f_{ss}$ that
develops is expected to be smaller than the force $f_{max}$ which
would be required to stretch a fully collapsed rod to the initial
length. This calculation yields a maximum force $f_{max}\approx10\,\mathrm{mgf}>f_{ss}$.

Lastly, Fig. \ref{fig:fixed_experiment}c depicts the total stretch
$\stretch$ and the local stretch in the collapsed ($\stretchg$)
and the swollen ($\stretchs$) domains as a function of time $t$.
The collapsed domains form very quickly and are typically associated
with a higher stretch. Once again, we find that the stretch $\stretchg$
is higher than $\stretchs$ for the same reasons as those described
in reference to Fig. \ref{fig:stretch_distribution}.

\subsection{A note on the influence of geometry on the coexistence of phases}

As mentioned by \citet{Suzuki&ishii99JCP}, geometry plays a significant
role in the presence of phase coexistence and VPTT. To understand
this, let us consider the initial (minimal) volume $v_{n}$ that is
associated with the onset of nucleation. Next, we examine the length
to diameter ratio $d/L$ of a dry cylindrical PNIPAM network subjected
to uniaxial extension around the VPTT. In the limit $d/L\ll1$, extension
leads to an initially stable nucleation of a swollen domain in the
rod with a volume $\sim v_{n}$. Further increase in tensile force
enlarges the volume fraction of the swollen domain until ultimately
the entire rod is in a swollen state. This case is the one considered
in this work. 

Alternatively, we can consider a dry PNIPAM disc with the dimensions
$d/L\approx1$ such that the total volume of the network $\pi d^{2}L/4<v_{n}$,
i.e. the minimum stable nucleation volume is larger than the volume
of the gel. Initial application of a uniaxial force leads to the extension
of the collapsed disc. However, once a critical uniaxial force is
applied, the globule-like chains are pulled out at once and an abrupt
and complete collapsed-to-swollen transition occurs. 

\subsection{A note on the influence of mechanical constraints on the VPTT}

The application of a mechanical force can be used to tune the VPTT.
The work of \citet{Suzuki&Sanda97JCP} demonstrated this with the
following experiment: a swollen PNIPAM rod was stretched from its
traction free state at $T=30^{\circ}C$ by a ratio $\stretch$ and
held fixed, where the range of investigated stretches was $\stretch=1-6$.
Next, the PNIPAM was subjected to a slow temperature increase in increments
of $0.05^{\circ}C$, and the volume and the force that developed were
measured. 

It was shown that the VPTT increased from $33.45^{\circ}C$ at $\stretch=1$
to slightly over $34.5^{\circ}C$ at $\stretch=6$. In addition, the
swollen-to-collapsed transition was accompanied by an increase in
the force required to maintain the constant length for the stretches
$\stretch=1-3.5$, while a decrease in force was measured for $\stretch>3.5$.
Interestingly, the volumetric deformations also depend on the stretch
and \citet{Suzuki&Sanda97JCP} demonstrated that the stress increases
upon the swollen-to-collapsed transition for all stretch values. 

The observations in that work can be explained through the local microstructural
evolution of the PNIPAM chains - at small fixed stretches $\stretch$
the forces on the chains are small and the energetic cost of the coil-to-globule
transition is low, but it is still higher than in a traction free
network. As the pre-stretch $\stretch$ increases, the chains experience
higher forces which prevent the local transitions and raise the VPTT.
While outside the scope of this contribution, these findings provide
a pathway to tune the VPTT through mechanical constraints. 

\section{Conclusions}

This contribution aims to explain the emergence of phase coexistence
in PNIPAM rods around the VPTT in the presence of mechanical constraints,
as demonstrated in the pioneering work of \citet{Suzuki&ishii99JCP}.
That paper reported two interesting experiments that led to non-trivial
behaviors: (1) the nucleation and time-dependent evolution of a collapsed
phase in a fixed swollen PNIPAM rod and (2) the stretch-induced formation
of swollen domains in a collapsed PNIPAM rod subjected to uniaxial
extension.

To delineate the mechanisms that enable phase coexistence, I began
by developing a local model for the PNIPAM chains. Chains below the
VPTT are hydrophilic with a long contour length whereas chains above
the VPTT are hydrophobic due to intramolecular interactions, which
lead to a shorter contour length. As seen in various experiments,
the application of a mechanical force to a globule-like chain leads
to the dissociation of the intramolecular interactions, resulting
in the ``pulling-out'' of chain segments that extend the contour
length and motivate water uptake. The stress associated with swollen
and collapsed networks was developed based on the local chain behavior.

Next, a method to describe the nucleation and evolution of a swollen
phase in a collapsed PNIPAM network subjected to uniaxial extension
was introduced. The local coil-to-globule transition is a stochastic
event, and accordingly a probabilistic approach was employed to determine
the state of a chain under a given mechanical force. The nucleation
and evolution of swollen domains requires the transition of many local
chains, and the likelihood of such an event was estimated from the
local probabilities of the chains. It is emphasized that phase coexistence
is maintained at equilibrium, and therefore the conditions for mechanical
and chemical equilibrium were summarized. 

To validate the proposed framework, I compared its predictions to
the two experiments reported by \citet{Suzuki&ishii99JCP}. The model
is capable of capturing the experimental findings and sheds light
on the microstructural evolution and local stress that develops during
the loading process.

The findings from this work show that mechanical constraints are the
main stabilizing source that enables simultaneous presence of collapsed
and swollen phases near the VPTT. Specifically, the deformation of
collapsed PNIPAM hydrogels at a temperature $T\sim\Tvptt$ can lead
to the pulling out of chains by breaking intramolecular bonds. Once
extended, the PNIPAM chains are hydrophilic and attract water molecules
in the environment, resulting in the nucleation of a swollen domain
that grows in size as additional chains open up as a result of an
increasing external force. It is important to point out that the stochasticity
of force-induced conformational transitions plays a key role in deciding
the position of the onset of nucleation and its growth. 

It was demonstrated that a similar effect occurs in constrained swollen
PNIPAM hydrogels that are heated to a temperature $T\sim\Tvptt$.
In this case, the chains in the network tend towards a globule-like
conformation. However, the external mechanical constraint hinders
and delays the transition, resulting in a time-dependent response
with a collapsed phase that increases in size until a steady state
is reached. 

Collectively, the findings from this work suggest a paradigm shift
in the control over the VPTT from traditional chemical design toward
mechanical force tuning. Specifically, as opposed to the classical
approaches, in which the VPTT is programmed through changes to the
chemical composition and network topology, this work sheds light and
provides guidelines on how mechanical constraints can be employed
to the same effect. Therefore, the proposed framework offers a pathway
to improve the control and performance of PNIPAM-based systems in
applications ranging from soft robotics to smart switches, where the
ability to precisely manipulate phase behavior is critical.

\bibliographystyle{biochem}
\bibliography{PNIPAM}

\end{document}